\documentclass[prb,aps]{revtex4}
\usepackage{graphicx}

\begin{document}

\title{Non-linear spin transport in a rectifying ferromagnet/semiconductor Schottky contact}
\author{R. Jansen, A. Spiesser, H. Saito and S. Yuasa}
\address{National Institute of Advanced Industrial Science
and Technology (AIST), Spintronics Research Center, Tsukuba,
Ibaraki 305-8568, Japan.}

\begin{abstract}
The electrical creation and detection of spin accumulation in ferromagnet/semiconductor
Schottky contacts that exhibit highly non-linear and rectifying electrical transport
is evaluated. If the spin accumulation in the semiconductor is small, the
expression for the spin voltage is identical to that of linear transport. However, if the spin
accumulation is comparable to the characteristic energy scale that governs the degree
of non-linearity, the spin detection sensitivity and the spin voltage are notably
reduced. Moreover, the non-linearity enhances the back-flow of spins into the
ferromagnet and its detrimental effect on the injected spin current, and the contact
resistance required to avoid back-flow is larger than for linear transport. It is also
shown that by virtue of the non-linearity, a non-magnetic metal contact can
be used to electrically detect spin accumulation in a semiconductor.
\end{abstract}

\maketitle

\section{Introduction}
\indent The development of electronic devices and circuits that use spin to encode digital
information is an attractive alternative to charge-based computing,
particularly if the unique attributes of semiconductors (amplification) and ferromagnets
(non-volatility) can be combined into a unified and energy-efficient computing technology.
Much progress has recently been made in developing the basic building blocks of such a
semiconductor spintronics technology \cite{chappert,awschalom,fertnobel,jansennmatreview,jansensstreview}.
Since mainstream semiconductors such as silicon
and GaAs are non-magnetic and in equilibrium do not possess any spin polarization, contacts
between a ferromagnet (FM) and a semiconductor (SC) are key components. On the one hand,
such junctions enable the transfer of spins from the ferromagnetic reservoir into the
semiconductor and thereby the creation of a sizeable (non-equilibrium) spin polarization of the
carriers in the semiconductor. Equally important, a ferromagnetic contact can
probe the spin polarization in a semiconductor and convert it into a detectable signal.\\
\indent In order to efficiently inject spins from a ferromagnetic metal into a semiconductor,
an interfacial energy barrier with a sufficiently large resistance times area (RA) product is
needed. This interface barrier limits the back-flow of the accumulated spins from the semiconductor
into the ferromagnetic source and avoids the so-called conductivity mismatch that prevents
spin injection from contacts with vanishingly small RA product \cite{jansensstreview,schmidt}.
Therefore, ferromagnetic contacts on semiconductors are either FM/I/SC structures, where
I is a thin tunnel insulator, or direct FM/SC contacts in which the Schottky barrier
formed at the metal/SC interface provides the energy barrier. In the latter case one
normally employs a semiconductor with a heavily-doped surface region or a $\delta$-doping
layer near the surface in order to obtain a Schottky barrier with a narrow depletion region
and an appropriate RA-product.\\
\indent Hitherto, it has been common practise to compare the experimental spin signals for
such structures to the theory previously developed for spin injection from a ferromagnetic
metal into a non-magnetic metal \cite{fertprb,fertieee,jaffres,maekawa,dery}, which starts
from a linear current-voltage relation. While this is appropriate for metallic junctions,
it does not capture the features that are specific for semiconductor junctions. These include
the energy band profile in the semiconductor and the associated energy barrier (Schottky
barrier), the localized states formed at the I/SC interface, as well as non-linear,
rectifying and/or thermally-activated transport. To better describe the experimental
results for magnetic tunnel devices on semiconductors, some of these aspects have
been examined \cite{derysham,jansenprl,tran,jansentwostep,derymr,tanamoto,selberherr}.
Notably, for FM/SC Schottky contacts it was described \cite{derysham} how spin transport is
changed due to the subsurface potential well that is formed in the semiconductor due to the
doping profile (heavily-doped surface layer on a substrate with lower doping density).
For FM/I/SC junctions, the presence of two barriers (tunnel insulator and Schottky barrier)
was shown to alter the spin detection efficiency when transport across the Schottky barrier
is by thermionic emission \cite{jansenprl}. Subsequently, spin injection by two-step tunneling via
interface states near the semiconductor surface was modeled and it was elucidated that this
transport process can modify spin signals in a profound way \cite{tran}. Important additions to
and refinements of the latter model have also been reported \cite{jansentwostep,derymr,note1}.\\
\indent Here we evaluate spin transport in a direct Schottky contact between a ferromagnetic
metal and a semiconductor with a homogeneous doping density. We consider electrical transport
across the interface that is highly non-linear (i.e., rectifying, diode-like) and present
a theory to describe the electrical creation of a spin accumulation and its electrical
detection via the Hanle effect. It is shown, by explicit evaluation starting from
spin-dependent non-linear transport equations, that for a rectifying Schottky
diode the expressions for the spin current, spin-detection sensitivity and the detectable
spin voltage signal are essentially the same as those of linear models, provided that the
spin splitting $\Delta\mu$ in the SC is small compared to the energy scale $E_0$ that governs
the degree of non-linearity of the transport across the interface. When $\Delta\mu$ is larger
and comparable to $E_0$, the non-linearity causes a reduction in the spin detection sensitivity
of the contact as well as a significant enhancement of the back-flow of spins into the
ferromagnetic electrode as compared to linear transport. Importantly, the non-linearity does not
produce any enhancement of the detectable spin signal. We discuss our results in light of
previous descriptions of the effect of the contact non-linearity on the spin
signals \cite{pu,shiogai,appelbaumdidv} in which the essential ratio of $\Delta\mu$ and $E_0$
does not appear. Finally, it is shown that the non-linearity enables a novel means to electrically
detect an (externally generated) spin accumulation in a semiconductor, namely by using
a {\em rectifying} contact with a {\em non-magnetic} metal electrode.

\section{Summary of linear spin transport theory}
\indent Let us first briefly summarize the results of the theory previously developed for
spin injection from a ferromagnetic metal into a non-magnetic metal, which starts from
a linear current-voltage relation \cite{fertprb,fertieee,jaffres,maekawa,dery}. The voltage
across such a contact is the sum of the regular resistive contribution ($R_0J$, with $J$ the
current density and $R_0$ the RA-product of the contact in the absence of spin accumulation)
and an additional contribution, the spin voltage, given by $P_G\,\Delta\mu /2$. Here
$P_G$ is the conductance spin polarization of the contact and $\Delta\mu = \mu^{\uparrow}-\mu^{\downarrow}$
is the induced spin accumulation, represented by a spin splitting between the electrochemical potentials
$\mu^{\uparrow}$ and $\mu^{\downarrow}$ of the electrons with spin pointing up or down, respectively.
The magnitude of $\Delta\mu$ is proportional to the density of injected spin current $J_s$
and to the so-called spin resistance $r_s$ of the non-magnetic material
(i.e., $\Delta\mu = 2\,J_s\,r_s$). Experimentally, the spin voltage can be detected \cite{jansennmatreview}
via a measurement of the Hanle effect, in which the spin accumulation is
reduced to zero by spin precession in an external transverse magnetic field. Keeping the
current constant, this results in a change in the voltage across the contact equal to the spin voltage.

\section{Theory of spin transport in a rectifying contact}
\indent The model we introduce to describe non-linear spin transport starts from
the expressions for electronic transport across a direct metal-semiconductor contact by
thermionic emission \cite{sze}. The basic parameters are the bias voltage $V$, the temperature $T$
and the height ${\Phi}_B$ of the Schottky barrier. Including the spin splitting $\Delta\mu$ of
the electrochemical potential in the semiconductor, the currents of majority and
minority spin electrons, $J^{\uparrow}$ and $J^{\downarrow}$, respectively, are:
\begin{eqnarray}
J^{\uparrow} = -J_0^{\uparrow}\left[exp\left(\frac{-q(V-\frac{\Delta\mu}{2})}{E_0}\right)-1\right], \label{eq1}\\
J^{\downarrow} = -J_0^{\downarrow}\left[exp\left(\frac{-q(V+\frac{\Delta\mu}{2})}{E_0}\right)-1\right]. \label{eq2}
\end{eqnarray}
Here, $q$ is the electronic charge and the voltage is defined as $V=V_{sc}-V_{fm}$, with $V_{fm}$ the
potential of the ferromagnetic electrode and $V_{sc}$ the spin-averaged potential of the semiconductor.
The forward bias condition (see Fig. 1) of the contact thus corresponds to negative voltage and current
density. The spin-dependent conductance of a Schottky contact with a ferromagnetic metal is included via the pre-factors:
\begin{eqnarray}
J_0^{\uparrow} = \left(\frac{1+P_G}{2}\right)A^{**}T^2\,exp\left(\frac{-q{\Phi}_B}{E_0}\right), \label{eq3}\\
J_0^{\downarrow} = \left(\frac{1-P_G}{2}\right)A^{**}T^2\,exp\left(\frac{-q{\Phi}_B}{E_0}\right). \label{eq4}
\end{eqnarray}
The $A^{**}$ is the modified Richardson's constant \cite{sze} that incorporates the finite probability of
reflection at the semiconductor-ferromagnet interface, which also produces the non-zero spin
polarization $P_G = (J_0^{\uparrow}-J_0^{\downarrow})/(J_0^{\uparrow}+J_0^{\downarrow})$ of the conductance
across the contact. The parameter $E_0$ is a characteristic energy scale that controls the degree of non-linearity.
For pure thermionic emission $E_0=kT$, with $k$ the Boltzmann constant. When there is also a contribution from
tunneling through the Schottky barrier the expressions have the same form, but the pre-factor is different and
the energy scale is changed to $E_0=nkT$, where $n$ is the so-called ideality factor \cite{sze}. Since $n\geq 1$
this reduces the non-linearity.\\
\indent At small enough bias ($|qV|<<E_0$) the transport approaches the linear regime without
rectification. Since spin transport in the linear regime has been described previously, we
shall focus here on the non-linear regime and consider a sufficiently large forward bias
such that $exp(-qV/E_0)>>1$. The total charge current $J= J^{\uparrow}+J^{\downarrow}$ and
the spin current $J_s= J^{\uparrow}-J^{\downarrow}$ across the contact are then:
\begin{eqnarray}
J = -exp\left(\frac{-qV}{E_0}\right)\left[J_0^{\uparrow}\, exp\left(\frac{+q\Delta\mu}{2E_0}\right)
+ J_0^{\downarrow}\, exp\left(\frac{-q\Delta\mu}{2E_0}\right)\right] \label{eq6} \\
J_s = -exp\left(\frac{-qV}{E_0}\right)\left[J_0^{\uparrow}\, exp\left(\frac{+q\Delta\mu}{2E_0}\right)
- J_0^{\downarrow}\, exp\left(\frac{-q\Delta\mu}{2E_0}\right)\right] \label{eq7}
\end{eqnarray}

\indent The way the spin accumulation is incorporated into the expressions for thermionic emission deserves some
attention. First, it is noted that the presence of the spin accumulation in the semiconductor does not affect
the current due to thermal emission of electrons over the Schottky barrier in the direction from the ferromagnet
to the semiconductor. The emission barrier height is given by the energy difference between the maximum of
the barrier and the electrochemical potential of the metal, both of which do not dependent on shifts of the
electrochemical potential in the semiconductor (see also Fig. 1). Hence, $\Delta\mu$ does not appear in the
second term between brackets in equations (\ref{eq1}) and (\ref{eq2}), just as the voltage does not appear,
for the same reasons \cite{sze}. In principle there is also a spin accumulation in the ferromagnetic metal,
but it is negligibly small owing to the very fast spin relaxation in ferromagnets. Secondly, in the theory of
thermionic emission transport \cite{sze} only the height of the Schottky barrier is relevant, not its shape.
Therefore, shifts of the electrochemical potential by a spin accumulation or by a voltage are equivalent \cite{note0}
and have the same effect on the thermionic emission current of electrons from the semiconductor to the ferromagnet.
The corresponding barrier heights for spin up and spin down electrons under forward bias ($V<0$ and $\Delta\mu<0$,
see also Fig. 1) are thus given by ${\Phi}_B+V-\Delta\mu/2$ and ${\Phi}_B+V+\Delta\mu/2$, respectively
(first term between brackets in equations (\ref{eq1}) and (\ref{eq2})).
\vspace*{5mm}

\begin{figure}[htb]
\hspace*{0mm}\includegraphics*[width=70mm]{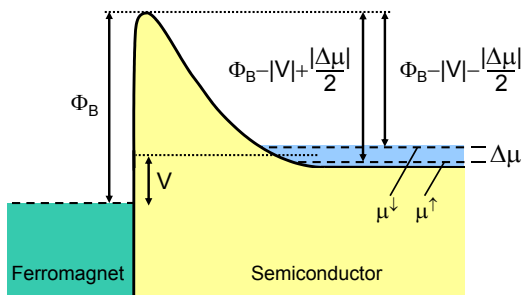}
\vspace*{0mm}\caption{Energy band diagram for a Schottky contact of a ferromagnetic metal on a semiconductor
under forward bias ($V<0$), corresponding to extraction of spins from the semiconductor. This produces a
negative spin accumulation ($\Delta\mu = \mu^{\uparrow}-\mu^{\downarrow} < 0$) in the semiconductor, as
indicated, assuming that the spin polarization $P_G$ of the conductance across the interface is positive.
The energy barriers for thermionic emission from the ferromagnet to the semiconductor (${\Phi}_B$) and from
the semiconductor to the ferromagnet for each spin (${\Phi}_B-|V|+|\Delta\mu|/2$ and ${\Phi}_B-|V|-|\Delta\mu|/2$,
respectively) are indicated.} \label{fig1}
\end{figure}

\subsection{Spin detection sensitivity and spin current}
\indent The existence of a spin accumulation can be detected electrically because the contact resistance
for forward bias depends on the value of $\Delta\mu$ (see eqn. (\ref{eq6})). The spin voltage signal
$\Delta V_{spin} = V(\Delta\mu) - V(\Delta\mu=0)$, obtained under the usual experimental condition
(Hanle effect measurement with the current kept constant \cite{jansennmatreview}), is given by:
\begin{equation}
\Delta V_{spin} = \frac{E_0}{q}\,ln\left[\left(\frac{1+P_G}{2}\right)\, exp\left(\frac{+q\Delta\mu}{2E_0}\right)
+ \left(\frac{1-P_G}{2}\right)\, exp\left(\frac{-q\Delta\mu}{2E_0}\right)\right] \label{eq10}
\end{equation}
This result captures the effect of the non-linearity: the spin voltage is not simply given by $P_G\,\Delta\mu /2$
and depends in a non-trivial manner on the magnitude of the spin accumulation.\\
\indent In the particular regime for which $|q\Delta\mu|<<2E_0$ we have
$\Delta V_{spin} = (E_0/q)\,ln\left[1 + P_G\,\left(q\Delta\mu/2E_0\right) \right]$, which, using
$ln(1+x)=x$ when $|x|<<1$, reduces to $\Delta V_{spin} = P_G\,\Delta\mu /2$.
This is exactly the same result as for the case of linear current-voltage characteristics. Hence, even for
highly non-linear and rectifying transport across a Schottky diode by thermionic emission, the spin detection
sensitivity of the contact ($\Delta V_{spin}/\Delta\mu$) is given by the linear response result, as long as
the magnitude of the induced spin splitting remains small compared to the characteristic energy scale
$E_0$ that parameterizes the degree of non-linearity (i.e., if $|q\Delta\mu|<<2E_0$). The spin detection
sensitivity as a function of $|q\Delta\mu|/E_0$ is depicted in the bottom panel of figure 2. Indeed, for
small $\Delta\mu$, the spin detection sensitivity is given by $P_G/2$. However, the spin detection sensitivity
decays when the value of the spin splitting becomes large and approaches $E_0$, and it even changes sign if
the spin splitting becomes much larger than $E_0$. These two features are not obtained in linear transport
models, for which the spin detection sensitivity does not depend on the magnitude of the spin accumulation.
We emphasize that the change of the spin detection sensitivity is a consequence of the non-linearity
of the transport, i.e., it is not related to the back-flow of spins into the ferromagnet (see below).
The non-linearity will also cause the Hanle line shape to deviate from the typical Lorentzian variation as a function
of magnetic field, because the spin detection sensitivity changes as the spin accumulation is reduced. This
aspect is not explored any further here. It is also noteworthy that for reverse bias ($V>0$), the spin detection
sensitivity goes to zero, since the reverse bias current is dominated by emission from the ferromagnet to the
semiconductor, which does not depend on $\Delta\mu$, as already mentioned.

\begin{figure}[htb]
\hspace*{0mm}\includegraphics*[width=85mm]{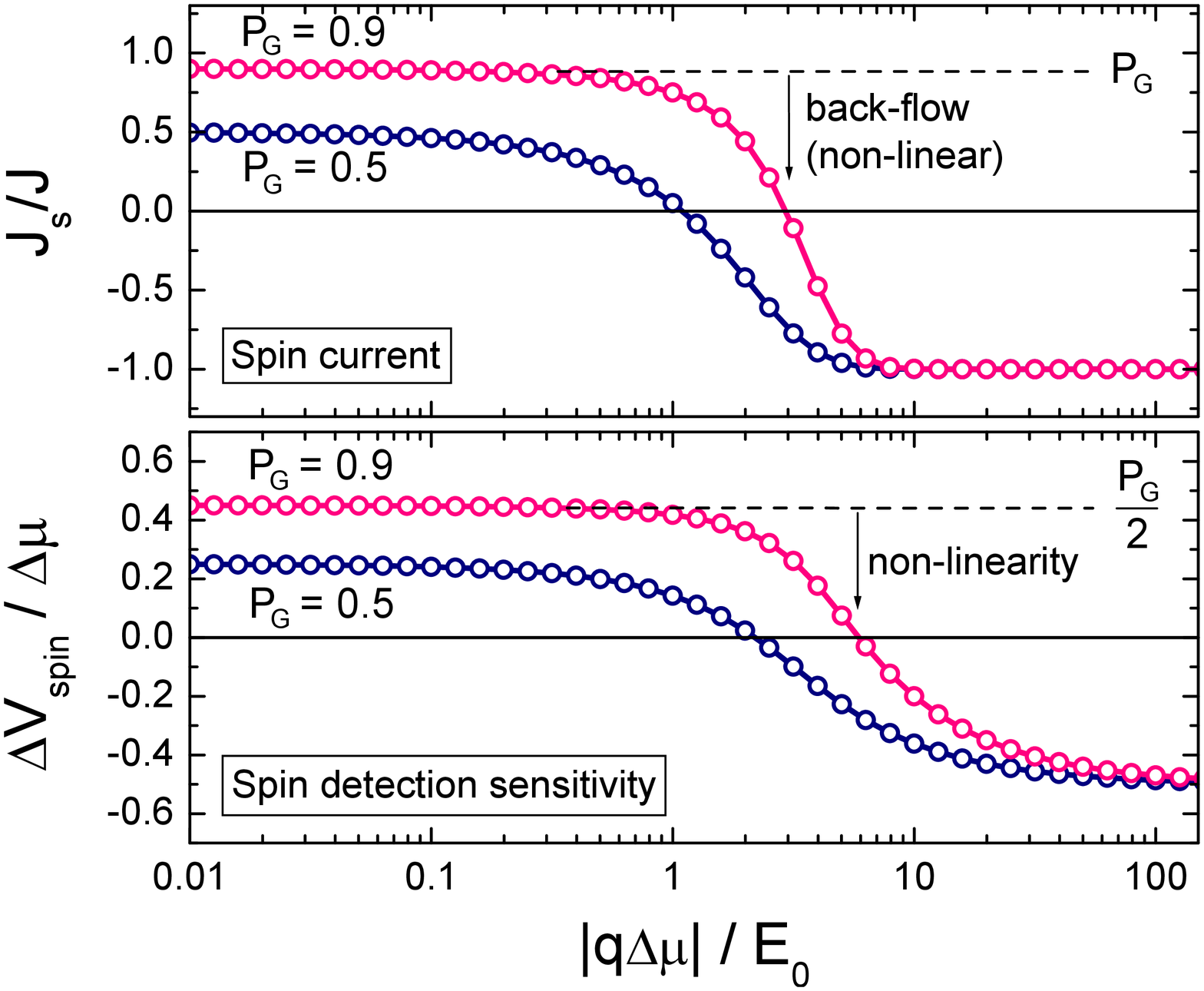}
\vspace*{0mm}\caption{Spin current ($J_s/J$, top panel) and spin detection sensitivity
($\Delta V_{spin}/\Delta\mu$, bottom panel) calculated for a rectifying Schottky contact of a
ferromagnetic metal and a semiconductor under sufficiently large forward bias (i.e., $exp(-qV/E_0)>>1$).
Results are shown as a function of the spin splitting $q\Delta\mu$ relative to $E_0$, for two different values
of the conductance spin polarization $P_G$. Since forward bias corresponds to the extraction of
spin-polarized electrons from the semiconductor and $P_G$ was taken to be positive, the induced
spin accumulation is negative, so the calculation was performed using negative values of
$\Delta\mu$. The regime for which the spin current is positive is relevant when the spin accumulation
is created by the contact itself, however, the regime for which the spin current is negative can only be
obtained when the spin accumulation is induced by an external source. The quantities on the
horizontal and vertical axis are all dimensionless.} \label{fig2}
\end{figure}

\indent Next, we evaluate the spin current density. From equations (\ref{eq6}) and (\ref{eq7}) we obtain:
\begin{equation}
J_s = J\,\left[\frac{J_0^{\uparrow}\, exp\left(\frac{+q\Delta\mu}{2E_0}\right)
- J_0^{\downarrow}\, exp\left(\frac{-q\Delta\mu}{2E_0}\right)}{J_0^{\uparrow}\, exp\left(\frac{+q\Delta\mu}{2E_0}\right)
+ J_0^{\downarrow}\, exp\left(\frac{-q\Delta\mu}{2E_0}\right)}\right] \label{eq13}
\end{equation}
In the limit $|q\Delta\mu|<<2E_0$ the spin current becomes:
\begin{equation}
J_s = J\,\left[\frac{P_G + \left(\frac{q\Delta\mu}{2E_0}\right)}{1 + P_G\left(\frac{q\Delta\mu}{2E_0}\right) }\right] \label{eq14}
\end{equation}
The spin current is shown as a function of $|q\Delta\mu|/E_0$ in the top panel of figure 2.
If the spin accumulation is small, the injected spin current is simply given by $P_G\,J$.
When the spin accumulation becomes larger and larger, the spin current is no longer independent
of $\Delta\mu$ and the existence of the spin accumulation reduces the injected spin current
(see Fig. 2, top panel). This phenomenon can be viewed as back-flow of spins into the
ferromagnetic electrode. Although this is well-established for linear
models \cite{fertprb,fertieee,jaffres,maekawa,dery} the parameters that control it are
different here because the back-flow is non-linear (see below).\\
\indent In principle, the spin current and the spin detection sensitivity of a ferromagnetic
Schottky contact can become negative due to the non-linearity (Fig. 2). However, an external (optical
or electrical) source of spins is needed to reach the regime with negative spin current and spin detection
sensitivity. If the same ferromagnetic contact is used to create and detect the spin accumulation, the detected
spin signal cannot change sign because the point where the spin detection sensitivity changes sign cannot
be reached. At very large density of injected current the spin accumulation becomes large, but during the
transient build up of the spin accumulation the injected spin current would first approach zero, and
beyond this point the spin accumulation does not increase any further. This saturation happens at a
value of the spin accumulation for which there is not yet a change in the sign of the spin detection
sensitivity (compare the zero-crossings in the bottom and top panels of Fig. 2).

\subsection{Spin accumulation and Hanle spin signal}
\indent The steady-state spin accumulation is obtained by defining the relation between spin accumulation,
injected spin current and the spin resistance $r_s$ of the non-magnetic material in the usual \cite{jansensstreview}
way ($\Delta\mu = 2\,J_s\,r_s$). If we insert expression (\ref{eq13}) for the spin current into
this we do not obtain an analytic solution, but we can solve for $\Delta\mu$ numerically. The resulting
Hanle spin signal, the so-called spin RA product $\Delta V_{spin}/J$, is shown as a function of the contact
RA product in Fig. 3. At large contact resistance, the injected spin current is small
and so is the induced spin accumulation. In this regime back-flow is negligible and the spin RA product is
equal to $P_G^2\,r_s$, which is identical to the result of linear transport models. As the junction RA product is reduced
below a certain value, the spin RA product decays. This is due to the back-flow of spins into the ferromagnet,
which becomes important for large $\Delta\mu$ as it limits the injected spin current, as already eluded to.
Although this type behavior is also obtained for linear transport (see the solid black curve in Fig. 3),
the point where back-flow starts to become relevant is different. For linear transport, back-flow is
significant when the contact RA product $R_0=V/J$ is smaller than the spin resistance $r_s$ of the
semiconductor. However, for the non-linear transport considered here, the point where back-flow sets in is
shifted to contact resistances significantly larger than $r_s$ (Fig. 3, symbols). Thus, the non-linearity
enhances the back-flow and the contact resistance needed to avoid it is larger than for linear transport.
The value of $\Delta\mu$ for which back-flow sets in is of the order of a mV for the parameters used here
($E_0=kT=25.8$ meV). These features make the non-linearity relevant for the design of devices such as
spin transistors with a FM source and drain contact, since these typically require large spin accumulation
and small contact resistance to obtain a large magnetic response and high speed operation.\\

\begin{figure}[htb]
\hspace*{0mm}\includegraphics*[width=78mm]{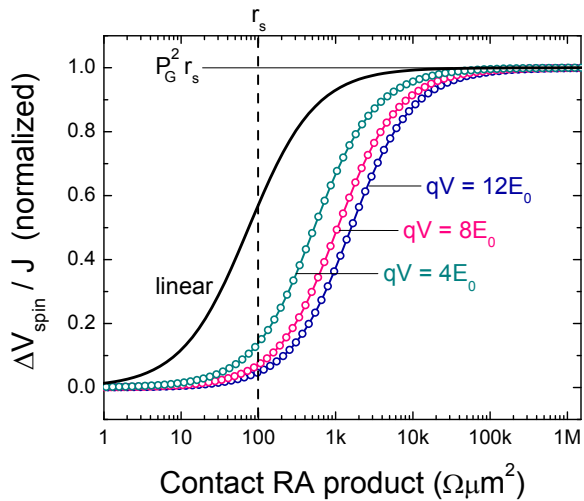}
\vspace*{0mm}\caption{Spin RA product ($\Delta V_{spin}/J$) versus contact RA product for a rectifying
Schottky contact of a ferromagnetic metal and a semiconductor under forward bias ($V<0$).
The contact resistance was varied via the Schottky barrier height ${\Phi}_B$. The spin RA product
is normalized to the value $P_G^2\,r_s$ obtained at large contact RA product. The parameters used are
$P_G=50\,\%$, $r_s=100\,\Omega\mu m^2$, T=300\,K and $V$ was varied from $-4kT/q$ to $-12kT/q$ (symbols).
The black solid curve is for linear transport.} \label{fig3}
\end{figure}

\indent We can gain some additional insight by using the approximate expression (\ref{eq14}) for the spin current
in the weakly non-linear regime, for which an analytic solution for the spin accumulation can be obtained:
\begin{equation}
\Delta\mu = 2\,P_G\,J\,r_s\,\left(\frac{R_0}{R_0 + \left(\frac{-qV}{E_0}\right)r_s}\right). \label{eq16}
\end{equation}
The term between brackets in equation (\ref{eq16}) describes the reduction of the spin accumulation due
to the back-flow. Whereas for linear transport this term is given by $R_0/(R_0+r_s)$, we find that
an additional factor $-qV/E_0$ appears. Since this factor is larger than unity for thermionic emission
in the forward bias regime under consideration, the effective spin resistance that controls the back-flow
is a factor of $-qV/E_0$ larger than $r_s$.

\subsection{Spin detection with a non-magnetic contact}
\indent A noteworthy aspect is that for a contact with a non-magnetic metal ($P_G=0$), the spin current
injected into the semiconductor is zero, but the spin-detection sensitivity is not (see figure 4). This is
readily understood. If transport across the contact is non-linear, then the change in current induced by
raising the electrochemical potential by $\Delta\mu /2$ for one spin direction is not compensated by the
change in current due to the lowering of the electrochemical potential by $\Delta\mu /2$ for the other spin.
The presence of a spin accumulation thus changes the total charge current across the contact. Hence, a
non-magnetic contact with strongly non-linear transport can be used to electrically detect a spin accumulation
(created by some external means, such as optically, or electrically from a nearby ferromagnetic injector).
Note, however, that for a non-magnetic contact the sign of the spin voltage $\Delta V_{spin}$ does not depend
on the sign of $\Delta\mu$, but is solely determined by the sign of the non-linearity (i.e., by the sign of
the second derivative $\partial ^2J/\partial V^2$). Also note that what we consider here is exclusively
due to the non-linearity of the transport {\em across} the detector contact interface. It should be distinguished
from effects due to the non-linearity of the transport {\em within the non-magnetic channel}, which in the
presence of a spin accumulation has been shown to lead to charge voltages that can be detected by a contact
with a non-magnetic electrode \cite{veramarun1,veramarun2}. This does not require nor rely on non-linear
transport across the interface between the channel and the non-magnetic electrode. Similarly, the detection
method we propose here does not require nor rely on non-linearity of the transport within the non-magnetic channel.

\begin{figure}[htb]
\hspace*{0mm}\includegraphics*[width=78mm]{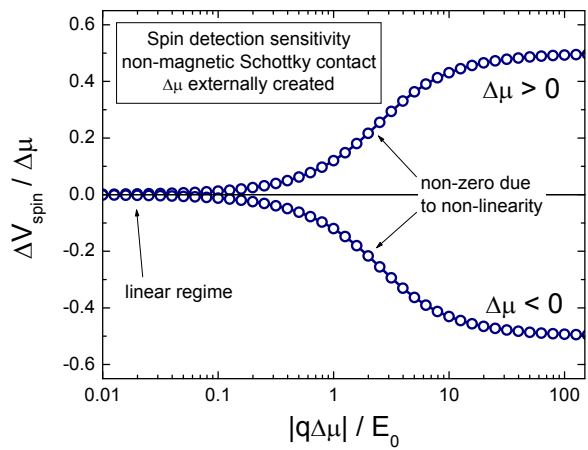}
\vspace*{0mm}\caption{Spin detection sensitivity ($\Delta V_{spin}/\Delta\mu$) calculated for a rectifying
Schottky contact of a {\em non-magnetic} metal and a semiconductor under forward bias ($V<0$).
The result is shown as a function of the spin splitting relative to $E_0$, for $P_G=0$. The spin accumulation
is created by an external source and can be positive or negative, which yields opposite signs of the spin detection
sensitivity, but note that the sign of the spin voltage $\Delta V_{spin}$ is always positive. The quantities
on the horizontal and vertical axis are dimensionless.} \label{fig4}
\end{figure}

\section{Discussion}
\indent Let us compare our results to previous works \cite{pu,shiogai,appelbaumdidv} that consider
the effect of the non-linearity on the spin signals. Those previous descriptions do not include the ratio
of $\Delta\mu$ and $E_0$ and thus do not capture the associated physics. When $\Delta\mu$ is small,
the transport parameters are essentially constant in the energy window defined by the spin accumulation,
and this feature must be reflected in the description of the effect of the non-linearity on the spin signals.
Another notable feature of previous works \cite{pu,shiogai} is the appearance of a multiplicative
factor $(\partial V/\partial J) / (V/J)$ in the expression for the spin voltage \cite{pu,shiogai},
suggesting that spin signal enhancement occurs if the differential resistance is larger than the regular
resistance $V/J$. Our results show that the effect of non-linearity on the magnitude of the spin signal
in general cannot be described by including this multiplicative factor (see also the appendix). In our
explicit evaluation the ratio of differential resistance and resistance does not appear, even though the transport by
thermionic emission is highly non-linear and rectifying \cite{note4}. Perhaps most strikingly, for reverse bias
the $\partial V/\partial J$ is much larger than $V/J$, but the spin signal is not enhanced, and in
fact, the spin detection sensitivity goes to zero at reverse bias, as already noted in section IIIA.
Another example that shows that the magnitude of the spin signal does not track $(\partial V/\partial J) / (V/J)$
is an Esaki diode. In the regime of negative differential resistance the Fermi level of the n-type
semiconductor is aligned with the band gap of the other (p-type) semiconductor. Creating a spin
splitting in the n-type semiconductor does not change the current because states around the Fermi
energy of the n-type electrode do not contribute to the current. Hence, the spin detection
sensitivity is zero. The differential resistance, however, is not zero. Hence, care should be taken
not to use the ratio of $(\partial V/\partial J)$ and $(V/J)$ in order to judge whether or not
the spin-detection sensitivity deviates from the linear result.\\
\indent Let us finally discuss whether the transport non-linearity can enhance the detectable spin signal
and thereby explain the results of experiments on spin transport in FM/I/SC tunnel devices. For various
semiconductors (GaAs \cite{tran}, Si \cite{jansennmatreview,jansensstreview,dash,jeon,uemura,dashschottky,sharmascaling},
Ge \cite{saitoge,jeonge,jain,ibage,ibagert,hanbickige,jaingeinterface,spiessergeox} as well as oxide
semiconductors \cite{jaffresoxide,parkinoxide,dashoxide}) these devices exhibit Hanle spin signals that are orders
of magnitude larger than what is expected from the theory previously developed for spin injection into
non-magnetic metals \cite{fertprb,fertieee,jaffres,maekawa,dery}. To explain the discrepancy, the role of
localized states in the tunnel oxide or at its interface with the semiconductor has been
considered \cite{tran,jansentwostep,derymr,uemura,casanova,casanova2}. In particular,
spin transport by two-step tunneling via localized interface states was modeled and predicted to yield
greatly enhanced spin signals due to spin accumulation in those interface states \cite{tran}. Because the
predictions of this model were shown to be inconsistent with experiments \cite{dash,jansensstreview,sharmascaling},
extended versions \cite{derymr,note1,casanova2} of the two-step tunneling model \cite{tran} have recently been developed.
Nevertheless, some of the predictions of those extended models \cite{derymr,casanova2} are also in
disagreement with experiments on electrical spin injection \cite{note2} and thermal spin injection by Seebeck
spin tunneling in similar FM/I/SC structures \cite{lebreton,jainthermal,jeonthermal,jeonvoltage}. The origin
of the large spin signals and whether localized states play a role is thus still unclear. With this
in mind, a recent experiment \cite{spiessermnge} has focused on a direct Schottky
contact of a metallic ferromagnet (Mn$_5$Ge$_3$) and a semiconductor (Ge), in which the absence of a
tunnel oxide eliminates all sources of spin signal enhancement that rely explicitly on localized states
associated with the oxide \cite{tran,jansentwostep,derymr,uemura,casanova,casanova2}. Nevertheless, the
observed spin signals \cite{spiessermnge}, that have all the characteristic features of spin accumulation
and spin precession due to the Hanle effect, were found to be up to 4 orders of magnitude larger than
predicted by linear transport models \cite{fertprb,fertieee,jaffres,maekawa,dery}. Since the studied
Mn$_5$Ge$_3$/Schottky contacts exhibited highly rectifying current-voltage characteristics, the question
arises whether the non-linear transport can affect the spin signal magnitude. Indeed, in some previous
reports it was argued that spin signals might be enhanced if transport is non-linear \cite{pu,shiogai}.\\
\indent Our explicit evaluation shows that a spin signal enhancement due to non-linearity is unlikely.
First of all, our results show that in the regime where non-linearity is important ($|q\Delta\mu|>E_0$), the
effect is to {\em reduce} the spin detection sensitivity, not to enhance it. In fact, it is straightforward
to show that non-linearity in general reduces the spin detection sensitivity because current-voltage
characteristics are typically super-linear (i.e., the conductance $(\partial J/\partial V)$ increases
with bias voltage). The spin detection sensitivity is enhanced only in special cases where transport
is sub-linear. Secondly, even if non-linearity is present, the induced spin accumulation is generally
small enough to ensure that $|q\Delta\mu|<<2E_0$, in which case the spin current, spin-detection
sensitivity and spin voltage signal are well described by the expressions previously derived for linear
transport. For instance, for strongly rectifying transport by thermionic emission we have $E_0=kT$, but since
$|q\Delta\mu|$ is typically a fraction of a meV in experiments conducted so far, the condition
$|q\Delta\mu|<<2E_0$ is satisfied at the temperatures used in the experiments. For transport with
weaker rectification, the value of $E_0$ is larger, and the non-linearity is even less likely to
play a role. Nevertheless, non-linearity might become important for technologically relevant devices in
which contacts with large spin polarization and very small RA product are used and the spin accumulation
becomes large.

\section{Summary}
\indent The theory presented here serves as a basis for the interpretation of spin transport in
rectifying ferromagnet/semiconductor Schottky contacts. It provides a quantitative means to assess
whether non-linear transport modifies the spin current, spin-detection efficiency and the detectable
spin voltage. The theory highlights the role of the magnitude of the induced spin splitting relative
to the energy scale that parameterizes the degree of non-linearity. If the spin accumulation is large
enough, the non-linearity is important, but it does not enhance the spin voltage. Rather, it reduces
the spin-detection sensitivity. It also enhances the back-flow of spins into the ferromagnetic injector
and its detrimental effect on the injected spin current. In order to suppress the back-flow, one needs a
larger contact resistance than what is deduced from linear transport models. These results are relevant
particularly for technological devices, since these generally require a large spin accumulation. It was
also shown that the non-linearity enables a novel means to detect a spin accumulation in a semiconductor,
using a non-linear contact with a non-magnetic metal electrode.

\begin{appendix}
\section{Effect of non-linear conductance on spin-detection sensitivity}
\indent In this appendix we discuss how the non-linear conductance of a ferromagnetic contact affects
its spin-detection sensitivity, and in particular we examine whether or not the spin-detection sensitivity
is modified by a factor $(\partial V/\partial J) / (V/J)$, as argued in previous works \cite{pu,shiogai}.
It is shown here that this multiplicative factor appears as a result of the (incorrect) assumption that the
non-linearity does not result in a change of the conductance when a spin accumulation is induced, but only
when the bias voltage changes.\\
\indent In order to illustrate this, we consider transport that, to first order, is linear in the voltage
$V$ and incorporate the non-linearity by using a conductance $G(V,\Delta\mu)$ that is a function of the
bias voltage {\em and} the spin accumulation. In the absence of a spin accumulation the voltage across
the contact is $V_0$ and the currents for each spin are simply given by:
\begin{eqnarray}
J^{\uparrow} = G_0 \left(\frac{1+P_G}{2}\right) V_0\\
J^{\downarrow} = G_0 \left(\frac{1-P_G}{2}\right) V_0,
\end{eqnarray}
where the $G_0$ denotes the conductance for $V=V_0$ and $\Delta\mu=0$. The total current $J$ is then $G_0\,V_0$.
In the presence of a non-zero spin accumulation, the voltage changes by an amount $\Delta V$ in order to keep
the total current unchanged, and the currents become:
\begin{eqnarray}
J^{\uparrow} = G(V,\Delta\mu) \left(\frac{1+P_G}{2}\right) \left(V_0 +\Delta V +\frac{\Delta\mu}{2}\right)\\
J^{\downarrow} = G(V,\Delta\mu) \left(\frac{1-P_G}{2}\right) \left(V_0 +\Delta V -\frac{\Delta\mu}{2}\right).
\end{eqnarray}
Importantly, the conductance $G(V,\Delta\mu)$ deviates from $G_0$ not only because the voltage has changed
by $\Delta V$, but also because the electrochemical potential in the non-magnetic electrode is shifted by
$\Delta\mu /2$ with respect to the average electrochemical potential (either up or down, depending on the
spin orientation). In general $\Delta V$ and $\Delta\mu$ are small compared to $V_0$, so that we can write:
\begin{eqnarray}
J^{\uparrow} = \left(G_0 + \frac{\partial G}{\partial V}\Delta V + \frac{\partial G}{\partial \mu}\frac{\Delta\mu}{2}\right) \left(\frac{1+P_G}{2}\right) \left(V_0 +\Delta V +\frac{\Delta\mu}{2}\right)\\
J^{\downarrow} = \left(G_0 + \frac{\partial G}{\partial V}\Delta V - \frac{\partial G}{\partial \mu}\frac{\Delta\mu}{2}\right) \left(\frac{1-P_G}{2}\right) \left(V_0 +\Delta V -\frac{\Delta\mu}{2}\right).
\end{eqnarray}
From the requirement that the total current with and without spin accumulation is the same and equal to $G_0\,V_0$,
and neglecting higher order terms proportional to $(\Delta V)^2$, $(\Delta\mu)^2$ or $\Delta V \Delta\mu$,
we obtain the voltage change $\Delta V$ as:
\begin{equation}
\Delta V = \left(\frac{P}{2}\right)\Delta\mu \left[\frac{G_0+\left(\frac{\partial G}{\partial \mu}\right) V_0}{G_0+\left(\frac{\partial G}{\partial V}\right) V_0}\right]
\end{equation}
The extra factor between the straight brackets represents the modification of the spin-detection sensitivity
due to the non-linearity of the transport across the contact. In order to obtain the change in the spin-detection
sensitivity, one thus needs to evaluate the derivatives $\partial G/ \partial V$ and $\partial G/ \partial \mu$,
which depend on the particulars of the transport across the contact. It is instructive to consider the
spin-detection sensitivity for two limiting cases:\\
If $\partial G/ \partial \mu=0$, then
\begin{equation}
\frac{\Delta V}{\Delta\mu} = \left(\frac{P}{2}\right)\left[\frac{G_0}{G_0+\left(\frac{\partial G}{\partial V}\right) V_0}\right] = \left(\frac{P}{2}\right) \left[\frac{\partial V/\partial J}{V/J}\right].
\end{equation}
If $\partial G/ \partial \mu= \partial G/ \partial V$, then
\begin{equation}
\frac{\Delta V}{\Delta\mu} = \left(\frac{P}{2}\right).
\end{equation}
We thus find that the multiplicative factor of $(\partial V/\partial J) / (V/J)$ discussed in previous
works \cite{pu,shiogai} appears as a consequence of setting $\partial G/ \partial \mu$ to zero.
In general this is not justified and $\partial G/ \partial V$ as well as $\partial G/ \partial \mu$
are non-zero. For instance, a change in voltage across an oxide tunnel barrier will change the energy
of the tunneling electrons with respect to the maximum of the potential barrier and thereby changes
the tunnel conductance. However, when a spin accumulation is created, the associated shifts of the
electrochemical potential (up or down depending on the spin) also change the energy of the tunneling
electrons with respect to the barrier maximum and thus the conductance. Similar statements can be
made for tunneling through a Schottky barrier, where the effective width and height of the barrier
change upon application of a voltage but also upon creation of a spin accumulation.
While this does not mean that $\partial G/ \partial V$ and $\partial G/ \partial \mu$ are identical,
the non-zero $\partial G/ \partial \mu$ counterbalances the effect of the non-zero $\partial G/ \partial V$.
This removes most of the multiplicative factor $(\partial V/\partial J) / (V/J)$ and yields a spin-detection
sensitivity that, despite the non-linearity, is close to the result $\Delta V / \Delta\mu = P/2$ obtained
for linear transport.\\
\indent Note that the above analysis applies to systems in which transport to first order is linear, such as
tunneling transport. For thermionic emission over an energy barrier, as discussed in the main
text, the transport equations are different, and this needs to be considered when evaluating the spin-detection
sensitivity \cite{note0}.

\end{appendix}


\begin{thebibliography}{99}
\bibitem{note0} The application of a bias voltage modifies not only the height of the barrier as seen by
electrons in the semiconductor, but also the barrier width, whereas the spin accumulation only changes
the height of the barrier by an amount $\pm q\Delta\mu/2$ depending on the spin. Nevertheless, shifting
the electrochemical potential by a spin accumulation or by a voltage has the same affect on the thermionic
emission current, because thermionic emission is well-described by a theory in which only the height of
the barrier is considered, and not its detailed shape.
\bibitem{note4} The nonlinearity of thermionic emission arises because transport is determined
by electrons in the high energy tail of the electron distribution, whose density depends
exponentially on energy. Since shifting the electrochemical potential by a spin accumulation
or by a voltage has the same affect on the thermionic emission current \cite{note0} it follows
that the spin detection sensitivity does not depend on the ratio of differential resistance
and resistance. When a spin accumulation is induced, the resulting current change is inversely
proportional to the differential resistance. In order to keep the current constant, the voltage
across the junction has to change, but this change is proportional to the same differential
resistance. Hence, the value of the differential resistance is immaterial.
\bibitem{note2} The predictions of the model by Song and Dery \cite{derymr} seem to be in disagreement
with existing experiments \cite{jansensstreview,dash,jeon,spiessergeox} in which no significant
reduction of the spin signal is observed when the bias voltage is changed from the regime with
$eV>kT$ to the regime with $eV<kT$. The current modulation that Song and Dery describe relies on
the condition $eV>kT$. Because of this condition, the relevant states in the emitting electrode
remain fully occupied, so that after an electron has tunneled for instance from the non-magnetic
electrode into a localized state in the barrier, the transport back into the non-magnetic electrode
is completely blocked. However, for $eV<kT$, the relevant states around the Fermi energy of the
emitting electrode are not fully occupied and transport back into the emitter is not blocked.
Consequently, the current modulation that Song and Dery describe is expected
to be reduced in a region having a width of a few $kT/e$ around zero bias voltage. Experimentally
this is not observed \cite{jansensstreview,dash,jeon,spiessergeox}. Moreover, because the model
by Song and Dery relies on the condition that $eV>kT$, it cannot account for the
obervation of thermally-driven spin injection without a charge tunnel current by Seebeck spin
tunneling \cite{lebreton,jainthermal,jeonthermal,jeonvoltage}.
\end{thebibliography}
\end{document}